\documentclass[aps,prl,reprint,groupedaddress,showpacs,amssymb,longbibliography,preprintnumbers]{revtex4-1}
\usepackage{graphicx}
\usepackage{hyperref}
\usepackage[normalem]{ulem}

\usepackage{color}
\definecolor{IITred}{rgb}{0.5,0.05,0.05}

\newcommand{\mev}{\ensuremath{\hbox{ MeV}}}
\newcommand{\gev}{\ensuremath{\hbox{ GeV}}}

\newcommand{\fm}{\ensuremath{\hbox{ fm}}}
\newcommand{\fb}{\ensuremath{\hbox{ fb}}}
\newcommand{\pb}{\ensuremath{\hbox{ pb}}}
\newcommand{\ifb}{\ensuremath{\fb^{-1}}}
\newcommand{\jpsi}{\ensuremath{J\!/\!\psi}}
\newcommand{\alphas}{\ensuremath{\alpha_{\mathrm{s}}}}

\newcommand{\orcid}[1]{\thanks{\href{http://orcid.org/#1}{ORCID: #1}}}
\newcommand{\cfrac}[2]{{\textstyle \frac{#1}{#2}}}
\newcommand{\half}{{\cfrac{1}{2}}}
\newcommand{\eq}[1]{Eq.~\ref{#1}}
\begin{document}
\preprint{FERMILAB--PUB--17/289--T}

\title{Heavy-Quark Symmetry Implies Stable Heavy Tetraquark Mesons $Q_iQ_j \bar q_k \bar q_l$}

\author{Estia J. Eichten}
\email[]{Email: eichten@fnal.gov}
\orcid{0000-0003-0532-2300}
\author{Chris Quigg}
\email[]{Email: quigg@fnal.gov}
\orcid{0000-0002-2728-2445}
\affiliation{Fermi National Accelerator Laboratory \\ P.O.\ Box 500, Batavia, Illinois 60510 USA}

\date{\today}

\begin{abstract}
For very heavy quarks $Q$, relations derived from heavy-quark symmetry predict the existence of novel narrow doubly heavy tetraquark states of the form $Q_iQ_j \bar q_k \bar q_l$ (subscripts label flavors), where $q$ designates a light quark. By evaluating finite-mass corrections, we predict that double-beauty states composed of $bb\bar u \bar d$, $bb\bar u \bar s$, and $bb\bar d \bar s$ will be stable against strong decays, whereas the double-charm states $cc \bar q_k \bar q_l$, mixed beauty+charm states $bc \bar q_k \bar q_l$, and heavier $bb \bar q_k \bar q_l$ states will dissociate into pairs of heavy-light mesons. Observation of a new double-beauty state through its weak decays would establish the existence of tetraquarks and illuminate the role of heavy color-antitriplet diquarks as hadron constituents.

\vspace*{6pt}
\noindent
\href{http://dx.doi.org/10.1103/PhysRevLett.119.202002}{DOI: 10.1103/PhysRevLett.119.202002}
\end{abstract}

\maketitle

Following the discovery of the charmonium-associated state $X(3872)$ by the BELLE collaboration~\cite{Choi:2003ue}, experiments have led a renaissance in hadron spectroscopy~\cite{[{For recent surveys of new states and candidate interpretations, see }][]Olsen:2014qna,*Lebed:2016hpi,*Esposito:2016noz,*Ali:2017jda}. 

Many of the newly observed states invite identification with compositions less spare than the traditional quark--antiquark meson and three-quark baryon schemes~\cite{[{The possibility of nonminimal configurations was foreseen in the foundational papers by }][{}]Zweig:1981pd,*Zweig:1964jf,*Gell-Mann:1964nj,*[{For an early survey of the emerging exotic spectroscopy, emphasizing the role of color couplings, see }][{}]Jaffe:2004ph}. Tetraquark states composed of a heavy quark and antiquark plus a light quark and antiquark have attracted much attention. The observed candidates all fit the form $c \bar c q_k \bar q_l$, where the light quarks $q$ may be $u, d, \hbox{or } s$. 
No such states are observed significantly below threshold for strong decays into two heavy-light meson states $\bar c q_k + c \bar q_l$; all have strong decays to $c \bar c$ charmonium + light mesons.  

In this Letter we examine the possibility of tetraquark configurations for which all strong decays are kinematically forbidden. We show that, in the heavy-quark limit, stable---hence exceedingly narrow---$Q_iQ_j \bar q_k \bar q_l$ mesons must exist. To apply this insight, we take into account  corrections for finite heavy-quark masses to deduce which tetraquark states containing $b$ or $c$ quarks should be stable. The most promising example is a $J^P=1^+$ isoscalar double-$b$ meson, $\mathcal{T}^{\{bb\}-}_{[\bar u \bar d]}$. 

In the heavy-quark limit, the lowest-lying tetraquark configurations resemble the helium atom, a factorized system with separate dynamics for the compact heavy color-$\mathbf{\bar 3}$  $Q_iQ_j$ ``nucleus'' and for the light quarks bound to the stationary color charge. (We recall that the one-gluon-exchange interaction is attractive for two quarks forming a color antitriplet, with half the strength of the attraction between a quark and antiquark bound in a color singlet.) At large $Q_i\hbox{ -- }Q_j$ separations, which become increasingly important as the heavy-quark masses decrease, the light $\bar q_k \bar q_l$  cloud screens the $Q_iQ_j$ interaction, so that the $Q_iQ_j \bar q_k \bar q_l$ complex may rearrange into a pair of heavy-light mesons~\cite{[{This behavior is exhibited in exploratory lattice QCD calculations reported in }][{}]Peters:2015tra,*Peters:2016isf,*[{It emerged in analytic calculations by }][{, who anticipated that, for infinitely heavy quarks $Q$, $QQ\bar{q}\bar{q}$ states would be stable against dissociation into two heavy-light mesons.}]Manohar:1992nd}. For heavy quarks $Q_iQ_j$ bound in a color $\mathbf{\bar 3}$ by an effective potential  of the ``Cornell'' Coulomb$\,+\,$linear form at half strength for both components~\cite{[{The strength of the Coulomb contribution is fixed by the color Casimir. Lattice studies indicate that the string tension for the color $\mathbf{\bar 3}$ is half that for the singlet configuration: }][{}]Nakamura:2005hk}, the rms core radii are $\langle r^2\rangle^{1/2} = 0.28\fm\, (cc); 0.24\fm\, (bc); 0.19\fm\, (bb)$, all considerably smaller than the size of the associated  tetraquark states.  Hence the core-plus-light (anti)quarks idealization should be a reliable guide to the masses of ground-state tetraquarks containing charms and bottoms.

The ground state of the attractive $\mathbf{\bar 3}$ $Q_iQ_j$ configuration may have total spin $S_{Q_iQ_j} = 1$ for identical quarks ($i=j$) or for quarks of different flavors ($i \ne j$) in a symmetric flavor configuration $\{Q_iQ_j\}$ or total spin $S_{Q_iQ_j} = 0$ for quarks of different flavors ($i \ne j$) in an antisymmetric flavor configuration $[Q_iQ_j]$. To construct a color-singlet $Q_iQ_j \bar q_k \bar q_l$ state, the light $\bar q_k \bar q_l$ must be in a color-$\mathbf{3}$. For the tetraquark ground state, both the heavy $Q_iQ_j$ and light $\bar q_k \bar q_l$ pairs must be in ($\ell = 0$) $s$-waves. To satisfy the Pauli principle, the flavor-symmetric $\{\bar q_k \bar q_l\}$ state must have total (light-quark) spin $j_\ell = 1$, whereas the flavor-antisymmetric $[\bar q_k \bar q_l]$ must have $j_\ell = 0$.

\emph{Stability in the heavy-quark limit.} For very heavy quarks, a hadron mass receives negligible contributions from the motion of the heavy quarks and spin interactions. Accordingly, the following relations hold among the masses of heavy-light and doubly-heavy-light mesons and baryons~\cite{[{The essence of this argument is developed in }][{}]Eichten:1987xu,*Lepage:1987gg,*[{For a comprehensive review, see }][{.}]Manohar:2000dt}:
\begin{widetext}
\begin{eqnarray}
   m(\{Q_iQ_j\} \{\bar q_k \bar q_l\}) - m(\{Q_iQ_j\}  q_y) &=& m(Q_x \{q_k q_l\}) - m(Q_x \bar q_y)   \nonumber \\
   m(\{Q_iQ_j\} [\bar q_k \bar q_l]) - m(\{Q_iQ_j\}  q_y) &=& m(Q_x [q_k q_l]) - m(Q_x \bar q_y)   \label{eq:hqs}\\ 
   m([Q_iQ_j] \{\bar q_k \bar q_l\}) - m([Q_iQ_j]  q_y) &=& m(Q_x \{q_k q_l\}) - m(Q_x \bar q_y)  \nonumber \\
   m([Q_iQ_j] [\bar q_k \bar q_l]) - m([Q_iQ_j]  q_y) &=& m(Q_x [q_k q_l]) - m(Q_x \bar q_y) \;. \nonumber  
  \end{eqnarray}  
\end{widetext}
(In the limit, a heavy core is a heavy core.) 

It is easy to see that the dissociation of $Q_iQ_j \bar q_k \bar q_l$ into two heavy-light mesons is kinematically forbidden, for sufficiently heavy quarks. The $\mathcal{Q}$ value for the decay is 
\begin{equation}
\begin{array}{r}
\mathcal{Q} \equiv m(Q_i Q_j \bar q_k \bar q_l) - [m(Q_i \bar q_k) + m(Q_j \bar q_l)] = \qquad  \\[3pt]
\Delta(q_k, q_l) - \half\!\left(\cfrac{2}{3}\alphas\right)^2\![1 + O(v^2)]\overline M + O(1/\overline M)\;,
\end{array}
\label{eq:twomesons}
\end{equation}
where $\Delta(q_k, q_l)$, the contribution due to light dynamics, becomes independent of the heavy-quark masses,  $\overline M \equiv (1/{m_Q}_i + 1/{m_Q}_j)^{-1}$ is the reduced mass of $Q_i$ and $Q_j$, and \alphas\ is the strong coupling. The velocity-dependent hyperfine corrections, here negligible, are calculable in the nonrelativistic QCD formalism~\cite{Caswell:1985ui}. For large enough values of $\overline M$, the middle term dominates, so the tetraquark is stable against decay into two heavy-light mesons.

The other possible decay channel is to a doubly heavy baryon and a light antibaryon,
\begin{equation}
(Q_iQ_j \bar q_k \bar q_l) \to (Q_iQ_j q_m) + (\bar q_k \bar q_l\bar q_m) \;.
\label{eq:baryons}
\end{equation}
By \eq{eq:hqs}, we have 
\begin{equation}
m(Q_iQ_j \bar q_k \bar q_l) - m(Q_iQ_j q_m) = m(Q_x q_k q_l) - m(Q_x \bar q_m) \;.
\label{eq:QvalB}
\end{equation}
In the heavy-quark regime, the flavored-baryon--flavored-meson mass difference on the right-hand side of \eq{eq:QvalB} has the generic form $\Delta_0 + \Delta_1/{M_Q}_x$. Using the observed mass differences, $m(\Lambda_c) - m(D) = 416.87\mev$ and $m(\Lambda_b) - m(B) = 340.26\mev$, and choosing effective quark masses $m_c \equiv m(\jpsi)/2 = 1.55\gev$, $m_b \equiv m(\Upsilon)/2 = 4.73\gev$, we find $\Delta_1 = 176.6\mev^2$ and $\Delta_0 =303\mev$,  hence the mass difference in the heavy-quark limit is $303\mev$. All of these mass differences are smaller than the mass of the lightest antibaryon, $m(\bar p) = 938.27\mev$, so we conclude that no decay to a doubly heavy baryon and a light antibaryon is kinematically allowed. \emph{This completes the demonstration that, in the heavy-quark limit, stable $Q_iQ_j \bar q_k \bar q_l$ mesons must exist.}

\emph{Beyond the heavy-quark limit.} To ascertain whether stable tetraquark mesons might be observed, we must estimate masses of the candidate configurations. Numerous model calculations exist in the literature~\cite{[{A useful compilation appears in Table IX of }][{}]Luo:2017eub}, but it is informative to  make estimates in the spirit of heavy-quark symmetry. 

The leading-order corrections for finite heavy-quark mass correspond to hyperfine spin-dependent terms and a kinetic energy shift that depends only on the light degrees of freedom,
\begin{equation}
\delta m  =   \mathcal{S}\frac{\vec{S}\cdot \vec{j_\ell}}{2 {\mathcal{M}}} + \frac{\mathcal{K}}{2{\mathcal{M}}} \;,
\label{eq:dm}
\end{equation}
where $\mathcal{M} = {m_Q}_i\mathrm{~or~}{m_Q}_i + {m_Q}_j$ denotes the mass of the heavy-quark core for hadrons containing one or two heavy quarks and  the coefficients $\mathcal{S}$  and $\mathcal{K}$ are to be determined from experimental data summarized in Table~\ref{tab:expmasses}. The spin splittings  lead directly to the coefficients $\mathcal{S}$ tabulated in the last column.
\begin{table*}[htbp]
    \caption{Representative masses~\cite{[{Except as noted, masses are taken from }][{ and 2017 update, \href{http://pdg.lbl.gov}{\texttt{pdg.lbl.gov}}.}]Olive:2016xmw}, in MeV, and derived quantities for ground-state hadrons containing heavy quarks.  \label{tab:expmasses}}
  \centering
      \begin{tabular}{@{} lcccccc @{}}
      \toprule
      State\footnote{Subscripts denote flavor-SU(3) representations for heavy baryons.} & $j_\ell$ & Mass $(j_\ell+\half)$ & Mass $(j_\ell - \half)$ & Centroid & Spin Splitting & $\mathcal{S}\hbox{ [GeV}^2]$ \\[2pt]
      \colrule
      $D^{(*)}$  $(c\bar d)$ & $\frac{1}{2}$ &$2010.26$ & $1869.59$ & $1975.09$& $140.7$ & 0.436  \\[0.2mm]
      $D_s^{(*)}$  $(c\bar s)$ & $\frac{1}{2}$ & $2112.1$ & $1968.28$ & $2076.15$& $143.8$ & 0.446  \\
      $\Lambda_c$ $(cud)_\mathbf{\bar{3}}$ & $0$ &$2286.46$ & $\cdots$ & $\cdots$ &  & $\cdots$ \\
      $\Sigma_c$ $(cud)_\mathbf{6}$ & $1$ & $2518.41$  & $2453.97$ & $2496.93$& $64.44$ & 0.132  \\
      $\Xi_c$ $(cus)_\mathbf{\bar{3}}$ & $0$ & $2467.87$ & $\cdots$ & $\cdots$ & & $\cdots$ \\
      $\Xi_c^\prime$ $(cus)_\mathbf{6}$ & $1$ & $2645.53$  & $2577.4$ & $2622.82$& $68.13$ & 0.141  \\
      $\Omega_c$ $(css)_\mathbf{6}$ & $1$ & $2765.9$  & $2695.2$  & $2742.33$& $70.7$ &  0.146 \\
      $\Xi_{cc}$ $(ccu)_\mathbf{\bar{3}}$ & $0$ & $3621.40$\footnote{From the LHC$b$ observation, Ref.~\cite{Aaij:2017ueg}.}  & $\cdots$ & & $\cdots$ \\[1mm]
      \colrule
      $B^{(*)}$ $(b\bar d)$ &$\frac{1}{2}$ & $5324.65$ & $5279.32$ & $5313.32$& $45.33$ & 0.427  \\[0.2mm]
      $B_s^{(*)}$ $(b\bar s)$ & $\frac{1}{2}$ & $5415.4$ & $5366.89$ & $5403.3$ & $48.5$ & 0.459  \\
      $\Lambda_b$ $(bud)_\mathbf{\bar{3}}$ &$0$ & $5619.58$  & $\cdots$ & & $\cdots$ \\
      $\Sigma_b$ $(bud)_\mathbf{6}$  &$1$ & $5832.1$ & $5811.3$ & $5825.2$& $20.8$ & 0.131  \\
      $\Xi_b$ $(bds)_\mathbf{\bar{3}}$ &$0$ & $5794.5$  & $\cdots$ & & $\cdots$ \\
      $\Xi_b^\prime$ $(bds)_\mathbf{6}$  &$1$ & $5955.33$ & $5935.02$ & $5948.56$&  $20.31$ & 0.128  \\
      $\Omega_b$ $(bss)_\mathbf{6}$  & $1$ & & $6046.1$ & & & \\[1mm]
      \colrule
      $B_c$ $(b\bar c)$ &$\frac{1}{2}$ & {6329}\footnote{Inferred from the lattice QCD calculation of Ref.~\cite{Dowdall:2012ab}.}  & $6274.9$  & {6315.4}$^\mathrm{c}$ & {54}$^\mathrm{c}$ & {0.340}$^\mathrm{c}$\\[1mm]
      \botrule
   \end{tabular}
\end{table*}
The pattern of the spin coefficients is entirely consistent with the expectations of heavy-quark symmetry.

The kinetic energy shift due to light quarks will be different in $Q\bar{q}$ mesons and $Qqq$ baryons. By comparing the centroid (or center-of-gravity, c.g.) masses for the charm and bottom systems we can extract 
the difference of the kinetic-energy coefficients $\mathcal{K}$ for states that contain one or two light quarks, viz.\ $\delta \mathcal{K} \equiv \mathcal{K}_{(ud)} - \mathcal{K}_d$.  For example, 
\begin{equation}
\begin{array}{r}
\{m[(cud)_\mathbf{\bar{3}}] - m(c\bar d)\} - \{m[(bud)_\mathbf{\bar{3}}] - m(b\bar d)\} \qquad  \\[3pt]
=  \delta \mathcal{K}\left(\displaystyle\frac{1}{2m_c} - \frac{1}{2m_b}\right) = 5.11\mev\;,
\end{array}
\label{eq:finda}
\end{equation}
from which we extract $\delta \mathcal{K} = 0.0235\gev^2$.  The resulting mass shifts are 
\begin{eqnarray}
\label{eq:KEshifts}
m[\{cc\} (\bar u\bar d)] - m(\{cc\}d)\!: &  & \frac{\delta \mathcal{K}}{4m_c} = 2.80\mev \\
m[(bc) (\bar u\bar d)] - m(\{bc\}d)\!: &  & \frac{\delta \mathcal{K}}{2(m_c+m_b)} = 1.87\mev \nonumber \\
m[\{bb\} (\bar u\bar d)] - m(\{bb\}d)\!: & \phantom{x} & \frac{\delta \mathcal{K}}{4m_b} = 1.24\mev \nonumber
\end{eqnarray}
These values are small---only slightly larger than the isospin breaking effects that we neglect as too small to affect the question of stability~\cite{[{Compare }][{}]Karliner:2017gml}.  

Combining the heavy-quark-symmetry relations of \eq{eq:hqs} with the leading-order corrections we obtain the masses of ground-state $Q_iQ_j \bar q_k \bar q_l$ tetraquarks summarized in Table~\ref{tab:2Q2q}~\footnote{Communication with decay channels tends to push the bound-state levels deeper. Open channels would induce mixing between the color-$\mathbf{\bar 3}$-core--$\mathbf{3}$-light quark configuration and meson--meson configurations.}. 
As inputs for the doubly heavy baryons not yet experimentally measured, we use the model calculations of Karliner and Rosner~\cite{[{}][{}]Karliner:2014gca}. 
\begin{table*}[htbp]
   \caption{Expectations for ground-state tetraquark masses, in MeV.\footnote{Masses of the unobserved doubly heavy baryons are taken from Ref.~\cite{Karliner:2014gca}; for lattice evaluations of $b$-baryon masses, see Ref.~\cite{Brown:2014ena}}  The column labeled ``HQS Relation'' is the result of our heavy-quark symmetry relations and is explicitly given by   the sum of the right-hand-side of \eq{eq:hqs} and the kinetic-energy mass shifts of \eq{eq:KEshifts}. Here $q$ denotes an up or down quark. For stable tetraquark states the $\mathcal{Q}$ value is highlighted in a box. \label{tab:2Q2q}}
   \centering
      \begin{tabular}{@{} lccccccc @{}}
\toprule
      State & $J^P$ &  $j_\ell$ & $m(Q_iQ_j q_m)$ (c.g.) & HQS relation &  $m(Q_iQ_j \bar q_k \bar q_l)$  & Decay Channel & $\mathcal{Q}$ [MeV] \\
\colrule
      $\{cc\}[\bar u \bar d]$ & $1^+$  & $0$ & $3663$\footnote{Based on the mass of the LHC$b$ $\Xi_{cc}^{++}$ candidate, $3621.40\mev$, Ref.~\cite{Aaij:2017ueg}.}  &$m(\{cc\}u) +  315$ &  $3978$  & $D^+{D}^{*0}$ 3876 & $102$ \\
      $\{cc\}[\bar q_k \bar s]$ & $1^+$  & $0$ & $3764$\footnote{Using the $s/d$ mass differences of the corresponding heavy-light mesons.}&$m(\{cc\}s) +  392$ &  $4156$  & $D^+{D}^{*-}_s$ $3977$ & $179$ \\
      $\{cc\}\{\bar q_k \bar q_l\}$ & $0^+,1^+,2^+$  & $1$ &  $3663$ &$m(\{cc\}u) +  526$  & $4146,4167,4210$  & $D^+{D^0}, D^+{D}^{*0}$ $3734, 3876$ & $412, 292, 476$\\
      $[bc][\bar u \bar d]$ & $0^+$  & $0$ & $6914$  & $m([bc]u) +  315$ &  $7229$  & $B^-D^+/B^0D^0$ $7146$& $83$\\
  $[bc][\bar q_k\bar s]$  & $0^+$ & 0 &  $7010$\footnote{Evaluated as $\half[m(c\bar s) - m(c\bar d) + m(b\bar s) -m(b\bar d)] + m(bcd)$.} & $m([bc]s) + 392$ & 7406 & $B_s D$ $7236$ & 170 \\     
     $[bc]\{\bar q_k \bar q_l\}$ & $1^+$  & $1$ & $6914$  & $m([bc]u) +  526$ &  $7439$  & $B^*D/BD^*$ $7190/7290$ & $249$ \\
 $\{bc\}[\bar u \bar d]$ & $1^+$  & $0$ & $6957$  & $m(\{bc\}u) +  315$ &  $7272$  & $B^*D/BD^*$ $7190/7290$& $82$\\
                     $\{bc\}[\bar q_k \bar s]$ & $1^+$ & 0 & $7053^{d}$ & $ m(\{bc\}s) + 392$ & 7445 &  $ DB_s^*$ 7282 & 163 \\
     $\{bc\}\{\bar q_k \bar q_l\}$ & $0^+,1^+,2^+$  & $1$ & $6957$  & $m(\{bc\}u) +  526$ &  $7461,7472,7493$  & $BD/B^*D$ $7146/7190$ & $317,282,349$\\
$\{bb\}[\bar u \bar d]$ & $1^+$  & $0$ & $10\,176$  & $m(\{bb\}u) +  306$ &  $10\,482$  & $B^-\bar{B}^{*0}$ $10\,603$& \fbox{$-121$} \\
      $\{bb\}[\bar q_k \bar s]$ & $1^+$  & $0$ & $10\,252^{\mathrm{c}}$  & $m(\{bb\}s) +  391$ &  $10\,643$  & $\bar{B}\bar{B}_s^*/\bar{B}_s\bar{B}^*$ $10\,695/10\,691$ & \fbox{$-48$} \\
      $\{bb\}\{\bar q_k \bar q_l\}$ & $0^+,1^+,2^+$  & $1$ & $10\,176$  & $m(\{bb\}u) +  512$  &  $10\,674,10\,681,10\,695$  & $B^-{B^0},B^-{B}^{*0}$ $10\,559, 10\,603$ & $115,78, 136$ \\
\botrule
   \end{tabular}
\end{table*}

\emph{Narrow Tetraquark States.}
    As we explained in the discussion surrounding \eq{eq:QvalB}, strong decays of $Q_iQ_j \bar q_k \bar q_l$ tetraquarks to a doubly heavy baryon and a light antibaryon are kinematically forbidden 
for all the ground states.   Strong decay to a pair of heavy-light mesons will occur if the tetraquark state lies above threshold.  For $J^P = 0^+\hbox{ or }2^+$,
a $Q_iQ_j \bar q_k \bar q_l$ meson might decay to a pair of heavy-light pseudoscalar mesons while for $J^P=1^+$ the allowed decay channel would be a pseudoscalar plus a vector meson. According to our mass estimates, the only tetraquark mesons below threshold are the axial vector $\{bb\}[\bar u \bar d]$  meson, $\mathcal{T}^{\{bb\}-}_{[\bar u \bar d]}$, that is bound by $121\mev$ and the axial vector $\{bb\}[\bar u \bar s]$ and $\{bb\}[\bar d \bar s]$ mesons bound by $48\mev$. 
We expect all the other $Q_iQ_j \bar q_k \bar q_l$ tetraquarks to lie at least $78\mev$  above the corresponding thresholds for strong decay~\cite{[{Note that if we took the SELEX value for the $\Xi_{cc}^{+}$ mass, $3519\mev$, rather than the LHC$b$ $\Xi_{cc}^{++}$ mass of Ref.~\cite{Aaij:2017ueg}, }][{, we would find $m(\{cc\}[\bar u \bar d])= 3876\mev$, coincident with the $3876\mev$ threshold for dissociation into a heavy-light pseudoscalar and heavy-light vector. Signatures for weak decay would include $D^+K^-\ell^+\nu$ and $\Xi_c^+\bar n$.}]Mattson:2002vu}. Promising final states  include $\mathcal{T}^{\{bb\}-}_{[\bar u \bar d]}\! \to \Xi^0_{bc}\bar{p}$, $B^-D^+\pi^-$, and $B^-D^+\ell^-\bar{\nu}$ (which establishes a weak decay), $\mathcal{T}^{\{bb\}-}_{[\bar u \bar s]}\! \to \Xi^0_{bc}\bar{\Sigma}^-$,  $\mathcal{T}^{\{bb\}0}_{[\bar d \bar s]}\! \to \Xi^0_{bc}(\bar{\Lambda},\bar{\Sigma}^0)$, and so on. 

As others have noted~\cite{Esposito:2013fma,Luo:2017eub}, unstable doubly heavy tetraquarks might be reconstructed as resonances in the ``wrong-sign'' combinations of $DD, DB,$ and $BB$. The doubly charged $\mathcal{T}^{\{cc\}++}_{[\bar d \bar s]} \!\to D^+ D_s^+, \hbox{ etc.}$ would stand out as \emph{prima facie} evidence for a non-$q\bar{q}$ level. 

While the production of $Q_iQ_j \bar q_k \bar q_l$ mesons is undoubtedly a rare event, we draw some encouragement for near-term searches from the large yield of $B_c$ mesons recorded in the LHC$b$ experiment~\cite{[{In 8-TeV $pp$ collisions, }][{ reported $8995 \pm 103$ $B_c$ candidates in $2\ifb$.}]Aaij:2014bva} and the not inconsiderable rate of Double-$\Upsilon$ production observed in 8-TeV $pp$ collisions by the CMS experiment, $\sigma(pp \to \Upsilon\Upsilon+\hbox{ anything}) = 68 \pm 15\pb$~\cite{Khachatryan:2016ydm}. The ultimate search instrument might be a future electron--positron Tera-$Z$ factory, for which the branching fractions~\cite{Olive:2016xmw} $Z \to b\bar{b} =15.12 \pm 0.05\%$ and $Z \to b\bar{b}b\bar{b} = (3.6 \pm 1.3) \times 10^{-4}$ offer hope of many events containing multiple heavy quarks.

\emph{Concluding remarks.} We have shown that, in the heavy-quark limit, stable $Q_iQ_j \bar q_k \bar q_l$ tetraquarks must exist. Our estimates of tetraquark masses  lead us to expect that strong decays of the $J^P = 1^+$ $\{bb\}[\bar u \bar d]$, $\{bb\}[\bar u \bar s]$, and $\{bb\}[\bar d \bar s]$ states  are kinematically forbidden, so that these states should be exceedingly narrow, decaying only through the charged-current weak interaction. Observation of any of these states would signal the existence of a new form of stable matter, in which the doubly heavy color-$\mathbf{\bar 3}$ $Q_iQ_j$ diquark is a basic building block.
The unstable $Q_iQ_j \bar q_k \bar q_l$ tetraquarks---particularly those with small $\mathcal{Q}$ values---may be observable as resonances decaying into pairs of heavy-light mesons, if they are not too broad to stand out above backgrounds.  

\begin{acknowledgments}
This manuscript has been authored by Fermi Research Alliance, LLC under Contract No.\ DE-AC02-07CH11359 with the U.S. Department of Energy, Office of Science, Office of High Energy Physics. The United States Government retains and the publisher, by accepting the article for publication, acknowledges that the United States Government retains a non-exclusive, paid-up, irrevocable, world-wide license to publish or reproduce the published form of this manuscript, or allow others to do so, for United States Government purposes.
\end{acknowledgments}

\noindent
\emph{Note added.}---We recently learned of interesting calculations of tetraquark masses that also highlight the likelihood of a stable doubly heavy tetraquark~\cite{[{}][{. This paper contains an extensive list of references to early work. }]Karliner:2017qjm,*Du:2012wp,*Chen:2013aba,*Francis:2016hui}. 

\bibliography{DHTQ}

\end{document}
%